\documentclass[conference]{IEEEtran}

\usepackage{cite}
\usepackage{graphicx}
\usepackage[hidelinks]{hyperref}
\usepackage{listings}
\usepackage{siunitx}
\usepackage{multirow}
\usepackage{xcolor}
\usepackage{xspace}
\usepackage{enumitem}
\usepackage{balance}
\usepackage{booktabs}
\usepackage[font=small,skip=3pt]{caption}
\usepackage{inconsolata}

\definecolor{backgroundColour}{HTML}{F8F8F8}
\definecolor{keywordclr}{HTML}{3969AC}
\definecolor{commentclr}{HTML}{A2A0AA}
\definecolor{stringsclr}{HTML}{11A579}
\definecolor{fnctionclr}{rgb}{0.467, 0, 0.533}
\definecolor{builtinclr}{rgb}{0.35, 0, 0.533}
\definecolor{symbolsclr}{rgb}{0.5, 0.25, 0.25}   
\definecolor{numbersclr}{rgb}{0.8, 0.2, 0}
\definecolor{bckgrndclr}{rgb}{0.91, 0.95, 0.95}
\lstdefinestyle{PythonStyle}{
    language=Python,
    backgroundcolor=\color{backgroundColour},
    keywordstyle=\color{keywordclr}\bfseries,
    stringstyle=\color{stringsclr},
    commentstyle=\color{commentclr},
    upquote=true,
    basicstyle=\ttfamily\linespread{0.9}\scriptsize,
    breakatwhitespace=false,
    breaklines=true,
    captionpos=b,
    keepspaces=true,
    numbers=left,
    numbersep=5pt,
    numberstyle=\color{commentclr}\ttfamily\tiny,
    showspaces=false,
    showstringspaces=false,
    showtabs=false,
    tabsize=2,
    xleftmargin=1.25em,
    frame=single,
    framexleftmargin=1.25em,
    morekeywords={assert,with,as,None,async,await}
}


\newcommand{\system}{\textsc{Academy}}

\begin{document}

\title{Empowering Scientific Workflows with\\Federated Agents}

\author{
   \IEEEauthorblockN{Alok Kamatar\IEEEauthorrefmark{1}}
   \IEEEauthorblockA{\textit{University of Chicago}}
   \\
   \IEEEauthorblockN{Mansi Sakarvadia}
    \IEEEauthorblockA{\textit{University of Chicago}}
   \and
    \IEEEauthorblockN{J. Gregory Pauloski\IEEEauthorrefmark{1}}
    \IEEEauthorblockA{\textit{University of Chicago}}
    \\
    \IEEEauthorblockN{Daniel Babnigg}
    \IEEEauthorblockA{\textit{University of Chicago}}
    \\~\\
    \and
    \IEEEauthorblockN{Yadu Babuji}
    \IEEEauthorblockA{\textit{University of Chicago}}
    \\
    \IEEEauthorblockN{Kyle Chard}
    \IEEEauthorblockA{\textit{University of Chicago}\\
    \textit{Argonne National Laboratory}}
    \\
    \and
    \IEEEauthorblockN{Ryan Chard}
    \IEEEauthorblockA{\textit{Argonne National Laboratory}}
    \\
    \IEEEauthorblockN{Ian Foster}
    \IEEEauthorblockA{\textit{University of Chicago}\\\textit{Argonne National Laboratory}
    }
}

\maketitle

\begingroup\renewcommand\thefootnote{\IEEEauthorrefmark{1}}
\footnotetext{Authors contributed equally to this work.}
\endgroup

\begin{abstract}
Agentic systems, in which diverse agents cooperate to tackle challenging problems, are exploding in popularity in the AI community.
However, existing agentic frameworks take a relatively narrow view of agents, apply a centralized model, and target conversational, cloud-native applications (e.g., LLM-based AI chatbots). 
In contrast, scientific applications require myriad agents be deployed and managed across diverse cyberinfrastructure. 
Here we introduce \system{}, a modular and extensible middleware designed to deploy autonomous agents across the federated research ecosystem, including HPC systems, experimental facilities, and data repositories.
To meet the demands of scientific computing, \system{} supports asynchronous execution, heterogeneous resources, high-throughput data flows, and dynamic resource availability.
It provides abstractions for expressing stateful agents, managing inter-agent coordination, and integrating computation with experimental control.
We present microbenchmark results that demonstrate high performance and scalability in HPC environments.
To explore the breadth of applications that can be supported by agentic workflow designs, we also present case studies in materials discovery, astronomy, decentralized learning, and information extraction in which agents are deployed across diverse HPC systems.
\end{abstract}

\begin{IEEEkeywords}
Computational Workflows, Distributed Computing, Federated Computing, Multi-Agent Systems, Open-Source Software
\end{IEEEkeywords}

\section{Introduction}
The desire to automate scientific processes has led to advancements in many fields, from artificial intelligence (AI)~\cite{zvyagin2023genslms} and computational workflows~\cite{deelman2009workflows} to research data management~\cite{bryce2012saasglobus} and self-driving laboratories (SDL)~\cite{abolhasani2023sdl}, but humans typically remain responsible for core aspects of the iterative research cycle, including hypothesis generation, experimental design, code development, and data analysis.
Automated components stall as humans must trigger experiments, manage workflows, correct errors, and make menial decisions.
This friction increases as the scale and ambition of computational science endeavors grow and leads to inefficient use of experimental and observational facilities, data repositories, and high-performance computing (HPC) systems.

Intelligent agents, either individually or composing larger multi-agent systems (MAS), rather than humans, can be the driving entities of discovery~\cite{pauloski2025agents}.
Agents are independent, persistent, stateful, and cooperative---working together to achieve a predefined goal with only intermittent human oversight.
The contemporaneous explosion of interest in multi-agent systems is largely a consequence of advancements in reasoning capabilities of the large language models (LLMs) often used to back AI agents~\cite{wei2022chain,wang2022self,sakarvadia2024towards}.
Expressing components of scientific applications as agents---programs that can perform tasks independently or semi-autonomously on behalf of a user or another agent---is powerful.
An agent manages its own local state and exposes a well-defined behavior.
Agents can perform human roles in iterative scientific processes~\cite{wang2023scientific} or encapsulate research cyberinfrastructure (e.g., computational resources and procedures, experimental instruments, data repositories)~\cite{gao2024agents}.

\begin{figure}
    \centering
    \includegraphics[width=\columnwidth]{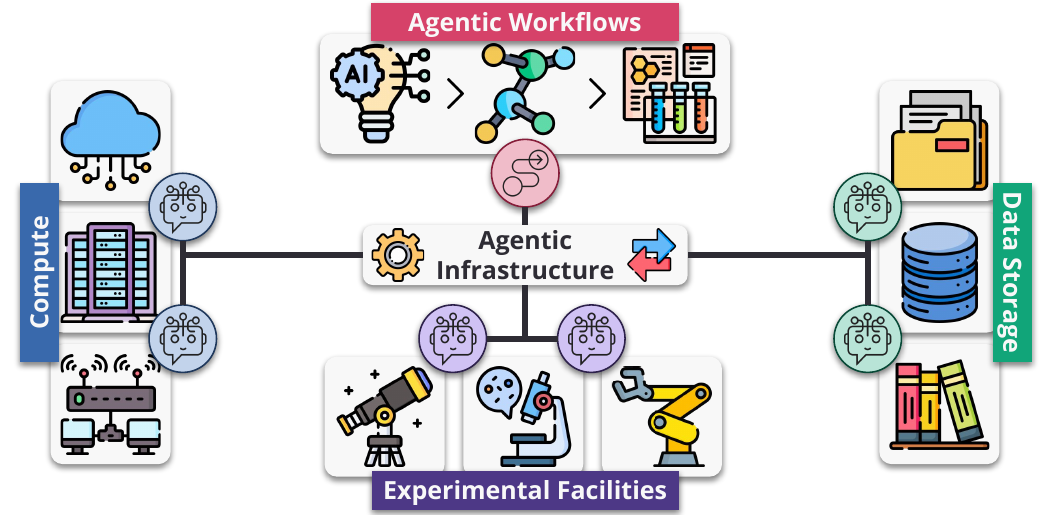}
    \caption{Cooperative agents, spanning federated research infrastructure (experimental facilities, computational systems, data storage), can enable agentic workflows that autonomously steer discovery.}
    \label{fig:ecosystem}
\end{figure}

However, current agent frameworks (e.g., AutoGen~\cite{wu2023autogen}) are not ready to build and deploy agents equipped for scientific applications.
They are limited in scope and typically apply a centralized model and target conversational, cloud-native applications (e.g., LLM-based AI chatbots)~\cite{langgraph,wu2023autogen,openai-swarm}.
Scientific applications, in contrast, typically
span geographically (and administratively) separated instruments, supercomputers, robotic facilities, and other resources that collectively constitute federated research infrastructure~\cite{osti2023iri}.
Federation creates unique challenges:
distributed resources have diverse access protocols, interactions between resources are asynchronous, and varying resource availability requires fault-tolerant and adaptive systems.
Existing agentic frameworks are not designed to address these intricacies and  workflow management tools designed for federated resources (e.g., Globus Compute~\cite{chard2020funcx}) cannot support the long-running and stateful properties of agents natively.
New middleware is needed to enable \emph{agentic workflows} that seamlessly integrate experiment, observation, simulation, analysis, and more, as in \autoref{fig:ecosystem}. 

We introduce \system{}, a novel actor-based framework for building agentic, scientific workflows, emphasizing modularity, statefulness, and interoperability across federated computing infrastructure.
Specifically, this work contributes:
\begin{itemize}[noitemsep,topsep=0pt,leftmargin=2em]
    \item \system{}, a novel, modular, and extensible middleware for expressing agentic workflows and deploying multi-agent systems across federated resources.
    \system{} addresses unique challenges in scientific applications, such as high data volumes, variable resource availability, and the heterogeneous nature of experimental and computational systems (\autoref{sec:design}).
    \item Performance analysis of \system{} in diverse scenarios yielding insights into the scalability and practical considerations of deploying agentic workflows (\autoref{sec:evaluation}).
    \item Case studies demonstrating the utility of agentic workflow design and highlighting improvements in automation, resource utilization, and discovery acceleration (\autoref{sec:case-studies}).
\end{itemize}
These contributions advance the state of the art in multi-agent systems for scientific discovery and establish a foundation for future innovations in autonomous research workflows.

\section{Background}
\label{sec:background}

Agents encompass a rapidly expanding front for AI research that can address a breadth of challenges across the computational sciences.
We begin with a definition of an agent---inspired by prior work---that is sufficiently generic to encompass the various semantic uses of the term.
Then, we describe applications that inform the design of \system{}, and distill requirements to support agentic scientific workflows.

An agent is a program that can perform actions independently or semi-autonomously on behalf of a user or another agent.
This definition incorporates both the modern notion of intelligent LLM agents~\cite{wu2023autogen} and the more traditional definitions of agents~\cite{goodwin1995formalizing, wooldridge1995agents, nwana1996agents, panait2005cooperative}.
While imprecise, it presents a powerful conceptual model for distributed computing. 
The agent concept originates from the actor model, a concurrent computing paradigm in which actors encapsulate a local state and communicate through message passing~\cite{hewitt1973actors}.
Agents extend the actor model with the notion of \emph{agency}---the ability of the agent to engage independently with its environment.

\subsection{Use Cases}
As scientists leverage more advanced computational and experimental resources, and pair them with increasingly capable machine learning models, such as LLMs, the nature of computational science workflows is changing. 
We highlight applications across four emerging patterns that are guiding our development of agentic middleware.

\paragraph{Steering Applications}
Scientists increasingly want to build applications that delegate the direction and composition of scientific campaigns to ML models.
Agents simplify the construction of these applications by observing and learning from the results of previous actions and adapting to meet the science goals.
For example, Colmena is an active learning library for steering simulation campaigns to discover molecules with desirable properties~\cite{ward2021colmena}.
In micro-genomics, scientists are using agents to learn the performance of viral identification tools and compose workflows based on sample characteristics~\cite{kolodisner2025viral}.
In physical simulations, agentic systems enable on-the-fly learning that dynamically switches between expensive subroutines and learned surrogate models based on the model uncertainty~\cite{tynes2025blend}.

\paragraph{Decomposing Applications across Facilities}
Multi-site workflows are becoming necessary to support complex and heterogeneous scientific applications.
These applications need to operate autonomously and manage local state while coordinating workflows across administrative domains.
MOFA~\cite{yan25mofa} is a materials science application designed to discover novel metal organic frameworks (MOFs) for carbon capture. It uses a chemical foundation model to generate candidate MOFs and several simulation tools to filter the candidates. The tools are best distributed across multiple computing facilities to meet the heterogeneous resource requirements. Agents can assist at managing the resources allocated at each facility. 
The Coalition for Epidemic Preparedness Innovations aims to shorten the timeline between disease outbreak and vaccine development to 100 days, which requires coordinating analytics across regions, continuously monitoring outbreaks and trial results, and integrating development and manufacturing into  workflows~\cite{saville2022cepi}. Agents can facilitate data, information and resource sharing, and automate components of the workflow.

\paragraph{Integrating Instruments into Workflows}
As experiments become more data-intensive, scientific applications integrate instruments into computational workflows, or analogously use ML models to control experimental facilities~\cite{vescovi2022linking}.
Embodied agents can integrate instruments into workflows and provide a mechanism to distribute control to the instrument or experimental site.
Integral field spectroscopy relies on precise data calibration to study distant spatially complex objects. By deploying agents with the instruments, scientists can track the optical parameters necessary for data-processing and accelerate discovery and steering of the telescope~\cite{babnigg2025ifum}. 
Self-Driving Laboratories (SDLs) provide an additional tool in chemistry and biology to automatically synthesize materials and measure properties~\cite{vescovi2023sdl}. Distributing agents to SDLs allows on-site control of experiments and integration of these experiments into computational workflows.

\paragraph{Conversational Research Assistants}
Scientists are building conversational assistants that help navigate literature, code, documentation, and other data.
These systems consist of multiple LLM-based agents that interact (e.g., via dialogue or shared context), with different agents being assigned different roles or capabilities.
For instance, Dr.\ MACS is an astronomy research assistant augmented with retrieval of astronomy literature~\cite{coburn2025drmacs}, and
ChemGraph is a general purpose chemistry assistant that delegates the construction of chemistry simulation workflows to an LLM~\cite{pham2025chemgraph}.
While constructed today with existing agent frameworks, these applications require access to scientific infrastructure to achieve the autonomy and performance required to meet scientific goals.

\subsection{Requirements to Support Federated Agents}
The case studies illuminate benefits of building agentic workflows in science, but also introduce new requirements unmet by existing agent, actor, or workflow systems.
 
\begin{enumerate}[label={(\bfseries R\arabic*)}, leftmargin=0.85cm]
    \item \textbf{Federated Orchestration.} Many of the use-cases require launching and managing agents across different computing systems or scientific instruments. 
    \item \textbf{Configurable Data Plane.} Agents may need to frequently communicate at the pace of simulation time-steps on high-performance interconnects or large volumes between facilities and instruments for long running campaigns. The communication mechanism that agents use needs to be configurable to adapt to the specific application.
    \item \textbf{Temporally Decoupled Messaging.} Research infrastructure has varying availability, typically with much lower uptimes than cloud infrastructure~\cite{papka2024alcfops}. Communication between agents must cope with facilities being temporarily unavailable.
    \item \textbf{Agent Authentication and Permissions.} Agents with the capability to use research infrastructure risk exposing the infrastructure to untrusted users.
    Researchers must be able to securely delegate (and revoke) the permissions to use tools and infrastructure to agents.
    \item \textbf{Resilient State Management.} State is a key feature across use-cases allowing these agentic systems to learn and adapt. While state is a necessary feature of the actor model, it is ill supported across workflow management tools that are popular among scientific applications.
\end{enumerate}

The adoption of specific frameworks is often also influenced by usability requirements such as intuitive and simple representations, local testing and debugging capabilities, etc.

\section{\system Design}
\label{sec:design}

In the design of \system{}, we aim to address the following high-level challenges: How to represent, in code, the declaration of and interaction between agents? How to deploy agents across federated infrastructure? How to achieve performance across heterogeneous systems and networks?
\system{} is an open-source Python library, available on GitHub\footnote{\url{https://github.com/academy-agents}} and PyPI.
We target Python for its broad compatibility with scientific workflow codes and libraries, but both the architecture and individual components could be implemented in other languages.

\subsection{\system{} Architecture}
\label{sec:design:architecture}

\system{} is a middleware for expressing agentic workflows and deploying multi-agent systems across federated resources.
Its architecture strongly decouples the implementation of agent behavior from execution and communication to simplify the development of new agents while maintaining flexibility in deployment.

\begin{figure}
    \centering
    \includegraphics[width=\columnwidth,trim=1mm 0 1mm 0,clip]{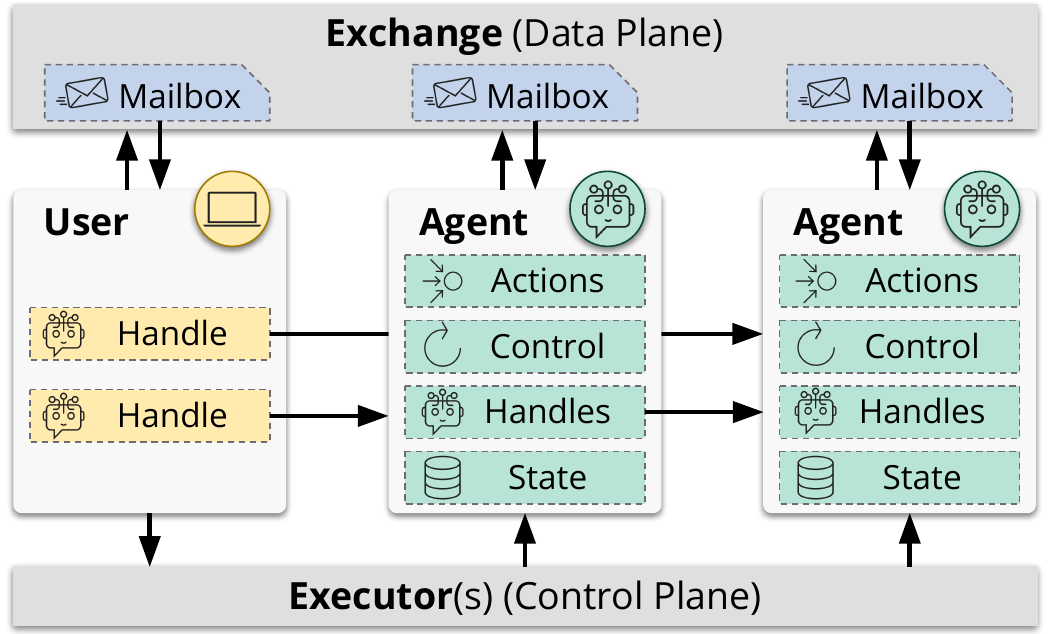}
    \caption{
    Agents and users in \system{} interact via handles to invoke actions asynchronously.
    Agents implement a behavior, defined by their actions, control loops, and state.
    \system{} decouples the control and data planes through the executor and exchange components that manage spawning agents and communication, respectively.
    }
    \label{fig:architecture}
\end{figure}

As depicted at a high level in \autoref{fig:architecture}, an \system{} deployment includes one or more \textit{agents} and zero or more \textit{users}.
An agent is a process with a local state, a set of actions, and a set of control loops.
Agents are executed remotely using an \textit{executor}.
Once running, an agent concurrently executes all of its control loops and listens for messages from clients, which can be other agents or users. 

A client interacts with an agent through a \textit{handle}, a term we borrow from actor frameworks.
A handle acts like a reference to the remote agent and translates method calls into action request messages.
Each entity (i.e., user or agent) has an associated \textit{mailbox} that maintains a queue of messages sent to that entity by other entities.
Mailboxes are maintained by an \textit{exchange} such that any client with access to a given exchange can send messages to the mailbox of another agent in the exchange and receive a response through its own mailbox.

\subsection{\system{} Interaction}
We first describe how scientists program agents and interact with the \system{} ecosystem.

\subsubsection{Agent Representation}
\begin{lstlisting}[
    style=PythonStyle,
    label={lst:behavior},
    caption={Example agent definition, showing initialization that sets an internal count variable, an action that squares a supplied value, and a loop that increments the internal count variable once a second.},
    float,
    floatplacement=t
]
from asyncio import Event, sleep
from academy.agent import Agent, action, loop

class Example(Agent):
    def __init__(self) -> None:
        self.count = 0  # State stored as attributes

    @action
    async def square(self, value: float) -> float:
        return value**2

    @loop
    async def count(self, shutdown: Event) -> None:
        while not shutdown.is_set():
            self.count += 1
            await sleep(1)
\end{lstlisting}
An agent is implemented as a Python class that inherits from the base \texttt{Agent} type: see \autoref{lst:behavior}.
This class-based approach is simple, so existing code can be easily transformed into agents, and extensible through inheritance and polymorphism.
Instance attributes maintain the agent's state, and methods define the actions and control loops.

The \texttt{@action} decorator marks a method as an action, allowing other entities to invoke it remotely.
An agent can invoke actions on itself, as actions are simply Python methods. Methods not decorated as \texttt{@action} are
private to the agent.
The \texttt{@loop} decorator marks methods as control loops.
Control loops extend the actor model to enable the programming of autonomous behavior.
Control loops are executed concurrently in an event loop; a shared \texttt{Event} is passed to each loop to signal agent shutdown,
so that control loops can exit gracefully.
A control loop can terminate early and the agent will remain running.
Commonly, control loops are used to execute a routine on a regular interval, such as to check the state of the environment, or in response to an event.
We provide special loop decorators, such as \texttt{@timer} and \texttt{@event}, that simplify agent implementations for common use cases.

Two special methods, \texttt{agent\_on\_startup()} and \texttt{agent\_on\_shutdown()}, allow agents to define callbacks when starting or shutting down, such as to load/store state or initialize/destroy resources.
Multiple inheritance of agents enables the creation of composite agents.

\subsubsection{Agent Invocation}
\label{sec:design:components:handle}
A \texttt{handle} is a client interface to a remote agent used to invoke actions, ping, and shutdown the agent. 
Each handle acts as a reference to that agent, translating each method call into a request message that is sent via the exchange and asynchronously waiting on the response message and returning the result.
The handle decides which mailbox to send from and listen to based on the context where the handle is used; a handle can be passed between users and agents and automatically attaches to the mailbox of the respective client.
This ensures that there is only one listening task per mailbox, and one mailbox per client (i.e., agent or user).

\subsubsection{Manager Class}
A \texttt{Manager} combines an exchange and one or more executors to provide a single interface for launching, using, and managing agents.
This reduces boilerplate code, improves communication efficiency, and ensures stateful resources and tasks are appropriately cleaned up.
An end-to-end example is provided in \autoref{lst:manager}.

\subsection{Agent Management}
\subsubsection{Agent Runtime}
\system{} provides an agent \texttt{Runtime} that executes an agent and manages communication with other entities.
It is instantiated with an agent class or instance, a unique identifier (the address of the agent's mailbox in the exchange), and an exchange interface.
When started, the \texttt{Runtime}: (1) invokes the \texttt{agent\_on\_startup()} callback of the agent, (2) spawns tasks for each \texttt{@loop} method, (3) spawns a task to listen for new messages in the agent's mailbox, and (4) waits for the agent to be shut down.
We use asyncio to ensure that many \texttt{@action} requests can be handled concurrently without blocking control loop and message listening tasks.

\subsubsection{Execution}
An agent can be run manually, but the intended method of deployment is via an \textit{executor}.
Any class that implements the \texttt{concurrent.futures.Executor} protocol can be used to launch agents on distributed resources via the \texttt{Manager} interface \textbf{(R1)}.
We use the following executors which cover most local and remote resource types:
\begin{itemize}[noitemsep,topsep=0pt,leftmargin=2em]
    \item \textbf{Thread:} Runs agents in separate threads of the same process. Useful for local development and testing or for lightweight or I/O bound agents.
    \item \textbf{Process:} Runs agents in distinct processes on a machine.
    \item \textbf{Parsl:} Runs agents across the workers of a Parsl Executor~\cite{babuji2019parsl}. Parsl supports execution on local, remote, and batch compute systems. 
    \item \textbf{Globus Compute:} Runs agents across Globus Compute Endpoints~\cite{chard2020funcx}. Globus Compute is a cloud-managed function-as-a-service (FaaS) platform which can execute Python functions across federated compute systems.
\end{itemize}

\subsubsection{State API}
The \system{} \texttt{State} API provides a dictionary-like interface 
for agents to persist in-memory state to storage.
This API currently supports only writing/reading state to/from disk, but could be extended to other storage modalities.   
Agents can define the \texttt{agent\_on\_startup()} callback to restore state automatically.
Given that research infrastructure can fail, agents may want to perform periodic state checkpointing \textbf{(R5)}.
\system{} does not enforce a specific checkpointing mechanism, as the format, location, and frequency of checkpoints are highly application specific, agents can use the \texttt{State} API to periodically store checkpoints.

\subsection{Agent Communication}
\label{sec:design:components:exchange}
Entities communicate by sending and receiving messages to and from mailboxes.
The mailboxes serve as mediated communication channels to decouple agents in time \textbf{(R3)}; messages persist in a mailbox even when an agent is offline.
An exchange hosts these mailboxes, and the \texttt{Exchange} protocol defines the interface to an exchange.
Namely, the \texttt{ExchangeClient} defines methods for registering new agent or user mailboxes, sending and receiving messages, and creating handles to remote agents.
Registering an agent or user involves creating a unique ID for the entity, which is also the address of its mailbox, and initializing that mailbox within the exchange.

A mailbox has two states: active and terminated.
\textit{Active} indicates that the entity is accepting messages, even if, for example, an agent has not yet started or is temporarily offline.
\textit{Terminated} indicates that the entity is permanently finished and will cause a terminated error to be raised by subsequent send or receive operations to that mailbox.

Users can define custom exchanges to address specific hardware or application characteristics \textbf{(R2)}.
We provide three exchange implementations: for testing (\textit{Thread Exchange}), single-site (\textit{Hybrid}), and distributed (\textit{Cloud}) deployments.

\subsubsection{Thread Exchange}
This implementation stores messages in-memory and is suitable for agents running in separate threads of a single process, such as when testing.

\subsubsection{Hybrid Exchange}
This implementation enables communication between entities across local networks (e.g., HPC interconnects).
It leverages an object store that persists information about registered entities.
A hybrid approach is used for message passing:
direct messaging is preferred, and indirect message passing via the object store is available as a fallback.
Upon startup, an entity writes its location (i.e., address and port) to the object store; peers that want to send a message can attempt to send directly to the entity's address.
If the peer is offline or a direct connection fails, 
messages are appended to the list of pending messages in the object store.
The multi-path design also makes the hybrid exchange suitable for dealing with heterogeneous networks.

Entities continuously listen to incoming messages from peers and pending messages in the object store.
Entities cache successful communication routes locally to reduce queries to the object store.
Our implementation uses TCP (Transmission Control Protocol) sockets for direct messaging and a Redis server as the object store.
Redis provides low-latency communication and optional replication, but applications that need greater fault-tolerance could consider (distributed hash table (DHT)-based object stores.
The security model for the hybrid exchange relies on Redis's built-in password authentication and access control lists.

\subsubsection{Cloud Exchange}
This implementation supports federated agent deployments. It exposes a secure, HTTP-accessible REST API that wraps Redis queues and enforces authorized access to mailboxes \textbf{(R4)}.

The cloud exchange security model is designed to ensure that any communication on the exchange is done by authenticated and permitted agents. 
This guarantees, for instance, that any action request served by an agent comes from an authorized agent/user and similarly response data is only read by the requesting agent/user.
Authorization and permissions enforcement occur at the exchange server; communication with the exchange is secured using standard TLS.
Users are authenticated via Globus Auth, an identity and access management platform that supports federated authentication via thousands of supported identity providers~\cite{tuecke2016auth}.
Each agent is registered as an authenticated entity with the Globus Auth service.
Creating an agent involves creating a new application identity with Globus Auth; a mailbox on the exchange linked with that identity, and a delegated token for the new agent to authenticate with the exchange; and launching the agent with the delegated token.

We assume that the launching entity is trusted, and that the launching mechanism is secure.
When launched, the new agent exchanges a delegated token for a refresh token, allowing long-lived access to the exchange without reauthentication. 
The refresh token is used to obtain access tokens to the exchange; these are introspected by the exchange to verify the client ID matches the expected client of the mailbox, and are cached for 60 seconds. 
Revocation of the delegated token invalidates the refresh token and any derived access token, but the exchange does not respond to the revocation until the cache expires. This response time depends on the cache length. 
Users can revoke tokens using the Globus Python SDK or web app.
Currently, permissions are coarse-grained---an agent is either allowed to communicate or not---but we plan to support finer-grained access controls for agents.

\subsubsection{Pass-by-Reference}
We optimize the Hybrid and Cloud exchanges for low latency, as control messages are typically small: $O$(100) bytes.
However, action request and response messages can contain arbitrarily sized serialized values for arguments and results that can induce considerable overheads when messages are sent indirectly via the object store.
To alleviate these overheads, we pass large values by reference and perform out-of-band data transfers by using ProxyStore~\cite{pauloski2023proxystore,pauloski2024proxystore}, which provides pass-by-reference semantics in distributed computing through proxy objects.
Proxy objects act like references (cheap to serialize and communicate) and automatically de-reference themselves to the true object using performant data storage and communication methods.
For example, ProxyStore can leverage RDMA (remote direct memory access) transfers via Mochi~\cite{ross2020mochi} and UCX~\cite{shamis2015ucx}, Globus Transfer~\cite{chard2014globus}, and reliable peer-to-peer UDP (user datagram protocol) through NAT hole-punching. 
Two key ProxyStore optimizations are useful within \system{}: proxies can be forwarded to actions executed on other agents without incurring additional data transfers and proxies can be asynchronously resolved to overlap communication and computation.

\begin{lstlisting}[
    style=PythonStyle,
    label={lst:manager},
    caption={Example of initialization, spawning, using, and shutting down an agent using the \texttt{Manager} interface.},
    float,
    floatplacement=t
]
from concurrent.futures import ThreadPoolExecutor
from academy.exchange.local import LocalExchangeFactory
from academy.manager import Manager

async with await Manager.from_exchange_factory(
    # Can be swapped with other implementations
    factory=LocalExchangeFactory(),
    executors=ThreadPoolExecutor(),
) as manager:
    agent = Example()  # From Listing 1
    handle = await manager.launch(agent)
    
    result = await handle.square(2)
    assert result == 4

    await handle.shutdown()  # Or via the manager
    await manager.shutdown(handle, blocking=True)
\end{lstlisting}

\section{Evaluation}
\label{sec:evaluation}

We studied the performance characteristics of \system{} to answer specific questions:
How well does the system scale?
How fast can agents be deployed?
What is the messaging latency?
In non-federated settings, we also compare to Dask and Ray, two popular frameworks with support for distributed actors in Python.

We conducted experiments using the Aurora supercomputer at the Argonne Leadership Computing Facility (ALCF), unless otherwise stated.
Aurora has \num{10624} nodes interconnected by an HPE Slingshot 11 network and a high performance DAOS storage system.
Aurora nodes contain two Intel Xeon Max CPUs, each with 52 physical cores and 64~GB of high-bandwidth memory; 512~GB of DDR5 memory per socket; and six 128~GB Intel Data Center Max GPUs (split into two tiles, or logical GPUs each).
In some cases we also use the Polaris supercomputer at ALCF and the \texttt{compute-zen-3} nodes of Chameleon
Cloud’s CHI@TACC cluster~\cite{keahey2020lessons}.
Polaris has \num{560} nodes interconnected by an HPE Slingshot 11 network.
Polaris nodes contains one AMD EPYC Milan
processor with 32 physical cores, 512 GB of DDR4 memory,
and four 40 GB NVIDIA A100 GPUs.
Each \texttt{compute-zen-3} node contains two 64-core CPUs and 256 GB memory.
Experiments were performed using Python 3.10, AutoGen 0.5.1, Dask 2025.2.0, Globus Compute 3.5.0, Parsl 2025.03.03, and Ray 2.43.0.

\subsection{Weak Scaling}
\label{sec:evaluation:scaling}

We measure weak scaling performance from two aspects: agent startup and action completion time.
We use the hybrid exchange with the object store located on the head node of the batch job to best match the behavior of Dask and Ray.

\begin{figure}
    \centering
    \includegraphics[width=\columnwidth]{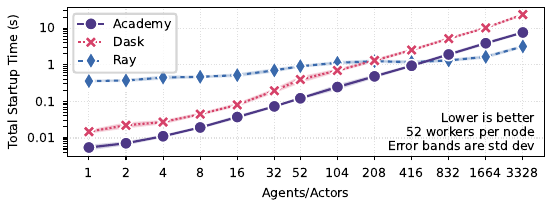}
    \includegraphics[width=\columnwidth]{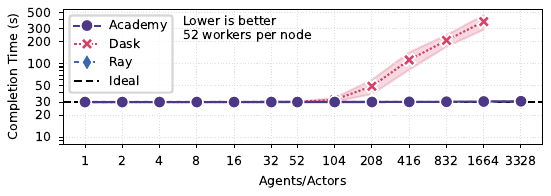}
    \caption{
    (Top) Warm-start times for $n$ agents/actors with \system{} (with Parsl launcher), Dask Actors, and Ray Actors.
    Ray does not benefit from warm starts because a new process is spawned for each actor.
    (Bottom) Time to execute 30 actions per agent/actor (weak scaling).
    Each action sleeps for 1~s.
    Note the \system{} and Ray lines are overlapped.
    }
    \label{fig:scaling}
\end{figure}

\subsubsection{Agent Startup Time}
We measure the time to spawn $n$ agents in \autoref{fig:scaling} (top).
We pre-warm the worker processes by starting and stopping $n$ agents, then record the average startup time over five runs.
Specifically, we measure the time between submitting the first agent to receiving a ping from all agents to ensure that they have finished their startup sequence.
We configured \system{} to use Parsl's High-throughput Executor as the executor.
Ray always spawns a new process per actor and thus does not benefit from pre-warmed workers leading to high startup overheads at smaller scales.
\system{}, Dask, and Ray have comparable cold start times, dominated by loading libraries from the shared file system.
With warm starts, \system{} starts a single actor in 5.5~ms, 2.8$\times$ faster than Dask.
\system{} scales well, starting \num{3328} actors in 7.6~s compared to Dask's 23.4~s, but Ray demonstrates an advantage at this scale with a 3.2~s startup.
Since \system{} can leverage many executor types, applications requiring frequent startup of agents can utilize Parsl for low-latency, and applications launching thousands of long-running agents could use Ray.

\subsubsection{Action Completion Time}
In \autoref{fig:scaling} (bottom), we execute 30 sleep tasks (1~s) per agent and record the total completion time.
We set the maximum concurrency to 1 for all agents to ensure that tasks are processed sequentially.
Completion time remains constant for \system{} and Ray up to 3328 agents while Dask performance degrades starting at 104 actors.

\subsection{\system{} Exchange}
\label{sec:evaluation:exchange}

\begin{figure}
    \centering
    \includegraphics[width=\columnwidth]{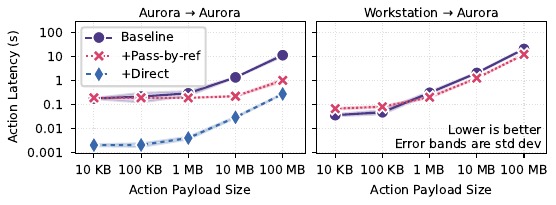}
    \caption{
    Time for a client to invoke a no-op action on an actor as a function of input and output payload size with different optimizations enabled on the hybrid and cloud exchanges.
    Two scenarios are considered: client and agent are at the same site using the hybrid exchange (left) and different sites using the cloud exchange (right).
    }
    \label{fig:exchange-ablation}
\end{figure}

Next, we investigate performance and optimizations of the hybrid and cloud exchange implementations.
In \emph{baseline}, all message data are communicated indirectly between peers via the exchange's object store.
For node to node communication on Aurora (Aurora → Aurora), the object store of the hybrid exchange is located on the head node of the Aurora batch job.
For remote communication, (Workstation → Aurora), the cloud exchange is deployed on AWS.
In \emph{pass-by-ref}, messages are still communicated via the object store, but action arguments and results are replaced with references using ProxyStore.
ProxyStore is configured to use ZeroMQ and P2P endpoints for intra-site and inter-site transfer of referenced objects, respectively.
In \emph{direct}, messages are communicated directly between peers, circumventing the object store; this is only possible when peers are located within the same site.

In \autoref{fig:exchange-ablation}, we measure the time it takes for a client to invoke a no-op action on an agent as a function of payload size.
We compare \textit{baseline}, \textit{pass-by-ref}, and \textit{direct} across two scenarios: \textit{Aurora → Aurora}, where the client and the agent are located on different Aurora nodes and messages are passed via the hybrid exchange, and \textit{Workstation → Aurora}, where the client is located on a personal workstation, the agent is on an Aurora node, and messages are passed via the cloud exchange.
The round trip network latencies are: Aurora compute node to Aurora head node: 0.23~ms; Aurora to AWS: 8.8~ms; Aurora to Workstation: 2.2~ms; and Workstation to AWS: 10.4~ms.

We observe that network latency to the exchange limits performance at smaller payload sizes ($\leq$ 100~KB).
\textit{Direct}, which is possible only in the intra-site scenario, circumvents these latencies.
In both scenarios, \textit{pass-by-ref} alleviates overheads of data transfer to and from the exchange by communicating data directly between the client and agent via ProxyStore.
For intra-site transfers, \textit{pass-by-ref} and \textit{direct} reduce action latency compared to the baseline by 91.2\% and 97.6\%, respectively, with 100~MB payloads.
For inter-site transfers, \textit{pass-by-ref} reduces action latency by 45.1\%.

\subsection{Agent Messaging}
Here, we investigate the performance of agent messaging.

\begin{figure}
    \centering
    \includegraphics[width=\columnwidth]{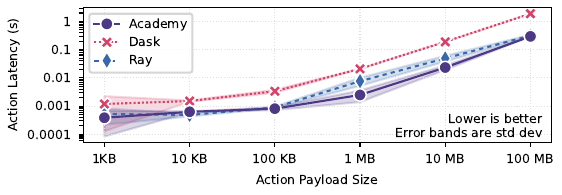}
    \includegraphics[width=\columnwidth,trim=1mm 0 1mm 0,clip]{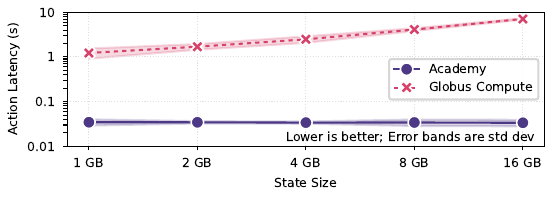}
    \includegraphics[width=\columnwidth]{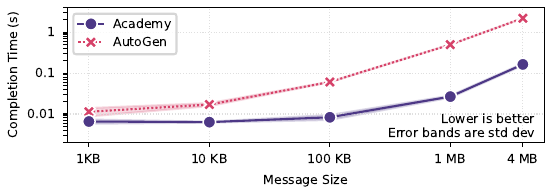}
    \caption{
    (Top)
    No-op action latency between two agents/actors running on separate Aurora nodes versus action input and output payload size.
    (Middle)
    Action latency between agents located on a workstation and Aurora. \system{} is compared to Globus Compute, a tool for building federated workflows that only support stateless functions, so state is read from the file-system on every invocation.
    (Bottom)
    Completion time for a simulated two agent chat where agents send ten messages back and forth with varied message sizes.
    \system{} is compared to AutoGen's distributed runtime.
    \label{fig:actions}
    }
\end{figure}

\subsubsection{Action Latency}
In \autoref{fig:actions} (top), we show action latency---the time between sending an action request and receiving a result---between two agents on different nodes.
We vary the input/output payload size to understand data transfer overheads.
The mean and standard deviation roundtrip latencies are 385$\pm$301~$\mu$s in \system{}, 1186$\pm$1059~$\mu$s in Dask, and 526$\pm$308~$\mu$s in Ray for the smallest 10~KB payloads, with latencies increasing with payload size.

For remote invocation, we compare the action latency of the cloud exchange to invoking functions with Globus Compute in \autoref{fig:actions} (middle).
Globus Compute is an alternative for building federated workflows, but only supports stateless functions.
To use state within Globus Compute, the state must be read from external storage (the shared file system) at every invocation, while when using \system{} the state remains in memory. 
The mean and standard deviation roundtrip latencies are 34$\pm$5 ms in \system{} and 1226$\pm$313 ms in Globus Compute for the smallest state size.
Dask and Ray do not support federated deployments---nodes in the cluster must have direct communication.

\subsubsection{Action Throughput}
We measure the maximum action throughput for a single agent by submitting a bag of no-op tasks to a pool of worker agents.
The pool contains 208 agents across four nodes to ensure that each worker agent is not over-saturated with work.
That is, the single submitter agent is the limiting factor for performance.
\system{}, Dask, and Ray achieve maximum throughputs of 3.4K, 185, and 14.1K action/s, respectively.
\system{} is 18$\times$ faster than Dask but 4$\times$ slower than Ray; however, this is a worst case scenario with no-op tasks and $>$3K actions/s is sufficient for real-world agents shown in \autoref{sec:case-studies}.

\subsubsection{Agent Conversations}
In \autoref{fig:actions} (bottom), we simulate a common pattern in LLM agents where two agents have a back-and-forth conversation.
We compare \system{} to AutoGen, a popular framework for creating multi-agent AI applications.
Each agent is run in a different process on the same node.
Agents send 10 messages back and forth, repeating with varying message sizes to simulate different kinds of conversations (i.e., text-only vs.\ multi-modal).
AutoGen's distributed agent runtime uses gRPC which has a maximum message size of 4~MB.
\system{} has comparatively lower overhead messaging in distributed settings.

\subsection{Memory Overhead}

\begin{figure}
    \centering
    \includegraphics[width=\columnwidth]{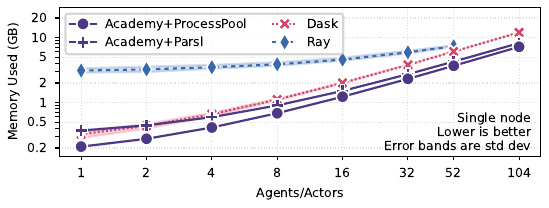}
    \caption{
    Memory used by $n$ agents/actors. We encountered a bug causing Ray to crash when deploying 104 actors on a single Aurora node (i.e., all cores on both sockets).
    }
    \label{fig:memory}
\end{figure}

We show memory used as a function of number of agents in \autoref{fig:memory}; for \system{},
we compare two executors: a low-overhead but single-node process-pool and Parsl's High-throughput Executor.
For fairness, we disable features in Dask and Ray that may increase memory, such as dashboards, and set the initial Ray object store size to the smallest possible value.
\system{} agents have low memory overheads, making them suitable for memory-constrained devices, as when deployed to edge devices via Globus Compute.
The Ray cluster head worker has high memory overhead, but that initial overhead is amortized as the number of actors is increased, indicating that each actor has modest overhead.

\section{Case Studies}
\label{sec:case-studies}

We use four applications to demonstrate the practicality, generality, and robustness of \system{} in real-world settings.
These examples illustrate how \system{} integrates with existing research infrastructure, enables distinct capabilities, 
and adapts to the varying demands of scientific applications.
More generally, these examples demonstrate how the agent paradigm simplifies the construction of the scientific application patterns identified in \autoref{sec:background}.

\subsection{Materials Discovery}
\begin{figure*}
    \centering
    \includegraphics[width=0.95\textwidth]{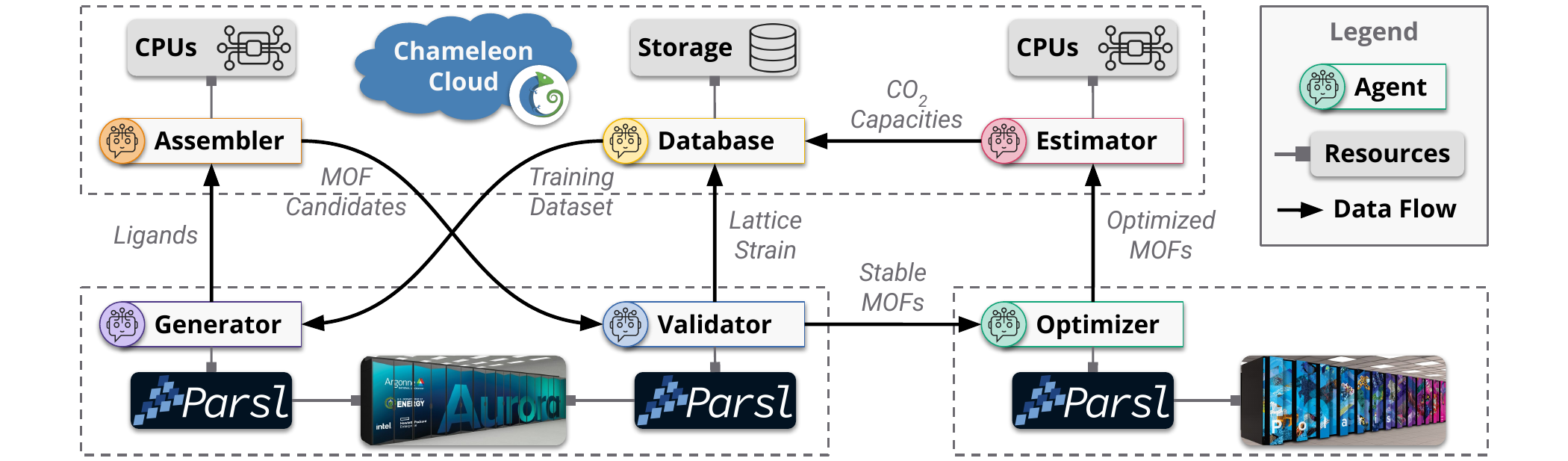}
    \caption{
    MOFA deploys agents across federated infrastructure with Globus Compute.
    The Assembler, Database, and Estimator run on Chameleon Cloud nodes with fast single-core performance; the Generator and Validator run on Aurora login nodes and execute AI and simulation tasks, respectively, on Aurora compute nodes; and the Optimizer runs on a Polaris login node and executes simulation tasks on Polaris compute nodes. 
    Each agent is responsible for a single MOFA stage. Agents cooperate by passing messages on the exchange, such as to request more work and trigger periodic events.
    Agents on Aurora and Polaris use Parsl to scale resources up and down based on workload needs.
    }
    \label{fig:app-mofa}
\end{figure*}

MOFA~\cite{yan25mofa} is an online learning application for generating, screening, and evaluating metal organic frameworks (MOFs) that couples generative AI methods with computational chemistry.
MOFs are polymers composed of inorganic metal clusters and organic ligands that are particularly suitable for gas adsorption applications such as carbon capture~\cite{furukawa2013chemistry}.
The goal of MOFA is to generate high-performing candidates by intelligently navigating 
space of possible MOF structures.
MOFA is representative of a broad class of scientific workflows that require careful integration of heterogeneous tasks spanning AI and simulation.

MOFA involves five stages:
(1)~a generative AI model produces candidate ligands;
(2)~these ligands are combined with predefined metal clusters to assemble candidate MOFs;
(3)~the candidates undergo iterative screening and validation using a series of molecular dynamics simulations;
(4)~CO\textsubscript{2} adsorption properties of the most promising structures are simulated and recorded in a database; and
(5)~the generative model is periodically retrained on the accumulated results to enhance its performance over time.

\begin{table}[]
    \centering
    \caption{Per-node throughputs, in MOFs/min, for the various components of the MOFA application (\autoref{fig:app-mofa}).}
    \label{tab:mofa-components}
    \begin{tabular}{rcrrcrr}
        \toprule
        Task && \multicolumn{2}{c}{Academy} && \multicolumn{2}{c}{Colmena/Parsl} \\
        && Machine & Throughput && Machine & Throughput \\
        \midrule
        Validate && Aurora & 10.73 && Polaris & 4.34  \\
        Optimize && Polaris & 0.85 && Polaris & 0.85 \\
        Estimate && Cloud & 6.81 && Polaris & 1.85 \\
        \bottomrule
    \end{tabular}
\end{table}

These stages have varied requirements that are best satisfied by
different computational resources (see \autoref{tab:mofa-components}).
The validation stage of the pipeline uses the LAMMPS GPU library to assess MOF stability (strain). As Aurora has more GPUs than Polaris (12 vs.\ 4), we can simulate MOFs approximately 4$\times$ faster on Aurora. Similarly for the CPU-only estimation stage, the the cloud node has 4$\times$ more CPU cores than a Polaris node allowing higher throughput screening. Meanwhile, the optimization stage uses CP2K, which could not be built on Aurora. Thus, to leverage the hardware best optimized for the specific computations, we must run the application across multiple sites. Furthermore, stateless execution, as supported for example by Globus Compute, is inadequate for this application. The generation component relies on a machine learning model that is costly to transfer and load in a stateless task. In addition, workflow components must adapt their resources autonomously to varying demands and scheduler conditions (e.g., job wall time limits or errors).

We adapt MOFA to use \system{} and deploy the resulting agentic application across federated resources: see \autoref{fig:app-mofa}.
We use six agents: Database, Generator, Assembler, Validator, Optimizer, and Estimator.
Each agent is responsible for a different component of the workflow and manages its own resources (i.e., storage and compute).
Agents are remotely deployed across Chameleon Cloud, Aurora and Polaris.
\begin{figure}
    \centering
    \includegraphics[width=\columnwidth,trim=0 1mm 0 1.2mm,clip]{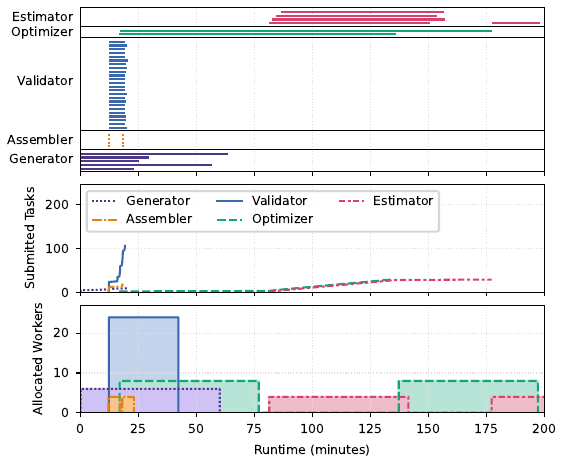}
    \caption{
    Execution trace of the Academy MOFA workflow.
    (Top) Active tasks per agent. 
    The vertical axis height represents the maximum size of the resource pool allocated by each agent.
    (Middle) Cumulative tasks submitted per agent.
    (Bottom) Active workers allocated in each agent's resource pool. 
    Worker allocations vary with demand (as in Assembler and Estimator) or batch job wall times (as in Generator, Validator, and Optimizer).
    }
    \label{fig:mofa-gantt}
\end{figure}

An execution trace of the agentic MOFA workflow (\autoref{fig:mofa-gantt})
shows how each agent scales out its allocated resources as work becomes available, and in the case of the Generator, Validator, and Assembler, scale down when their workload decreases.
After a ramp up period, the Optimizer consistently has work to do but their batch jobs within which workers run have 60 minute wall times that expire and then must be resubmitted, causing temporary drops in the number of workers.
Active tasks that are killed are restarted in the next job.
This separation of concerns is key to enabling long-running workflows---resource infrastructure is not persistently available and agents will need to be able to adapt to that varying availability.

\begin{figure}
    \centering
    \includegraphics[width=\columnwidth,trim=1mm 1mm 1mm 1.2mm,clip]{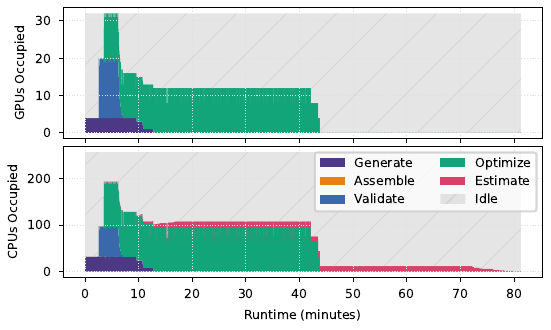}
    \caption{
    Resources used for a traditional workflow implementation of MOFA deployed on Polaris.
    The static resource configuration does not match the workload, so in the estimation stage, when tasks are only using the CPU, the rest of the resources are idle.
    }
    \label{fig:mofa-colmena}
\end{figure}

We contrast the Academy implementation of MOFA to a version implemented with Colmena~\cite{ward2021colmena},
a bespoke simulation campaign framework built on Parsl~\cite{babuji2019parsl}---a traditional workflow system that does not support stateful federated agents. 
We compare the Academy version, deployed on federated resources, to the Colmena version run as a single batch job on Polaris. 
In addition to using different resources, 
the Academy execution includes queue wait times to acquire resources while the Colmena execution does not.
Although the performances are thus not directly comparable, the comparison against Colmena/Parsl (shown in \autoref{fig:mofa-colmena}) demonstrates how federated agents support more efficient resource utilization. With Academy, resource allocations (\autoref{fig:mofa-gantt}) are managed autonomously by each agent, whereas Colmena allocates resources statically for the entire workflow. 
This, along with the hardware configuration of Polaris, means that when the Colmena workflow performs CPU-only tasks like ``estimate,'' the GPU resources of the allocation sit idle.

A further advantage is that the loose coupling between agents in the Academy model makes it trivial to swap one agent implementation for another, provided that the API exposed by the agent remains the same.
It also becomes easier to integrate future agents, such as to incorporate embodied agents that interact with self-driving labs to synthesize the best-performing MOFs via physical experiments~\cite{abolhasani2023sdl,vescovi2023sdl,zhao2025artificial}.

\subsection{Astronomical Spectroscopy}
Integral field spectroscopy is a powerful technique used in observational astrophysics to gather spatially-resolved spectroscopic data from a target field at once, enabling the study of spatially-complex objects like distant strongly-lensed galaxies---optimal candidates for understanding galaxy formation in the early universe~\cite{mateo2022ifum}.
Before analyzing the spectra offered by the Integral Field Units for Magellan (IFU-M) instrument, scientists first process and calibrate the data which involves the overarching steps: pre-processing, spatial and wavelength calibrations, and datacube creation.
A crucial step in pre-processing the data is detecting and filtering cosmic rays through subtracting multiple images of the same patch of sky.
However, the optics of the telescope shift over time in response to small variations in external factors such as temperature.
The images must be aligned by finding offset and scaling parameters that minimize the variation between images.
Previously, the alignment was found through a grid search of parameters between pairs of images.

\begin{figure}
    \centering
    \includegraphics[width=\columnwidth,trim=0 1mm 0 1.2mm,clip]{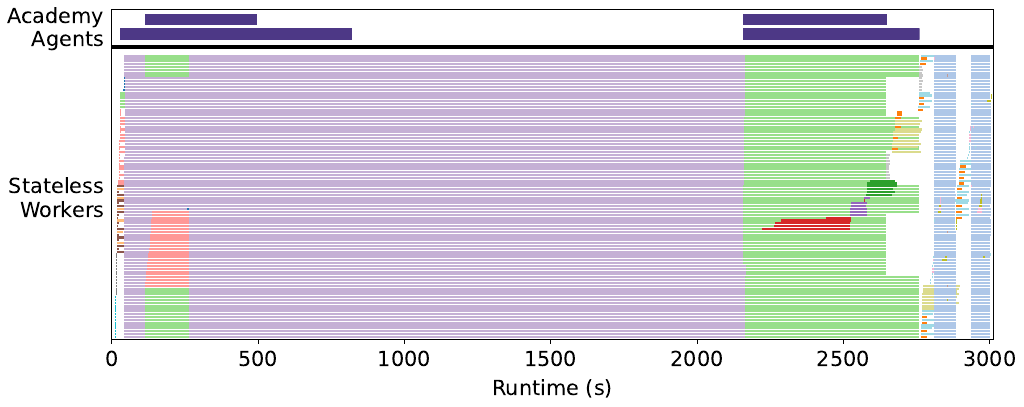}
    \caption{Execution trace of the IFU-M workflow. Each color corresponds to a different task type. Tasks within the Parsl workflow are stateless, but invoke actions on agents that track the instrument state.}
    \label{fig:app-ifu}
\end{figure}

We employ a hybrid pattern for this workflow in which
Academy conceptually maps the instrument into the workflow by estimating and storing the optical parameters of the telescope (see \autoref{fig:app-ifu}).
As these parameters vary smoothly in time, the agent uses previously calibrated images to inform the search for new parameters, leading to faster alignment and reduced noise in the processed images.
The two spectrographs of the instrument are each represented by their own agent.
Other tasks in the application remain stateless, allowing them to be scheduled to any available worker.
In the future, deploying the agents on the instrument would allow the agent to use observations (e.g., temperature measurements) to inform the alignment, and provide on-site feedback for targeting observations.

\subsection{Decentralized Learning}
In decentralized machine learning a set of models learn collaboratively across distributed datasets~\cite{hegedHus2019gossip}. This paradigm is particularly relevant today as data are generated in decentralized settings and transfer to a centralized location can be infeasible for cost and privacy reasons.
Each device in a decentralized learning workflow performs three steps: 
(1)~train model on local data for a set number of iterations;
(2)~receive models from neighboring devices and send its own model to neighbors; and
(3)~update the local model via an all-reduce operation performed across its own and received models.
It is straightforward to reframe such a decentralized learning workflow as an agentic workflow. 
We represent the application as a graph in which nodes are agents and edges are communication channels. 
Each agent trains its local model, receives neighboring agents' models, and periodically aggregates received models with its own model. 

\begin{figure}
    \centering
    \includegraphics[width=\columnwidth,trim=1.2mm 1.2mm 1.2mm 1.2mm,clip]{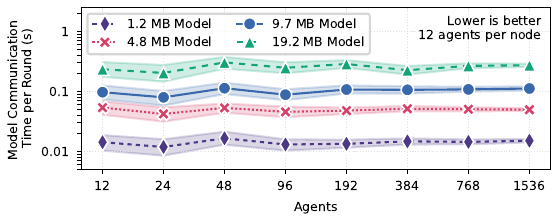}
    \caption{
    Model communication time to an agent's neighbors averaged
    over five rounds of decentralized training. Training time and aggregation time are excluded since they are nearly constant.
    }
    \label{fig:app-distributdML}
\end{figure}

Using \system, we simulate decentralized learning on Aurora. For the connectivity between devices, we choose a power-law cluster graph to approximate real-world networks~\cite{holme2002growing}.
Each agent uses a copy of the MNIST dataset~\cite{deng2012mnist}. The agents are configured to use \textit{pass-by-ref} with ProxyStore as the transfer backend.
We investigate the cost of distributing updates from all agents as we scale the size of the graph for different model sizes: see \autoref{fig:app-distributdML}. 
The agents are deployed on Aurora using Parsl, with each agent pinned to a single GPU tile (two tiles per physical GPU), allowing 12 agents per node. 
Our results demonstrate \system{}'s ability to support more than 1500 autonomous agents working collaboratively with no client coordination, as shown by the constant times seen in \autoref{fig:app-distributdML}.

\subsection{Information Extraction}

\begin{figure}
    \centering
    \includegraphics[width=\columnwidth,trim=1.2mm 1.2mm 1.2mm 1.2mm,clip]{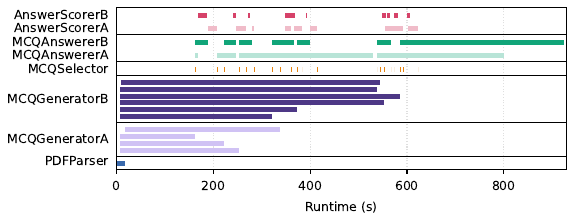}
    \caption{
    Execution trace of the agentic MCQ workflow processing 10 manuscripts to generate and validate questions and answers over 15 minutes. The figure shows the active agents and the duration of their tasks. Agents employ either the Mistral-7B-Instruct-v0.3 or Meta-Llama-3-70B-Instruct model, denoted A and B, respectively.
    }
    \label{fig:mcq}
\end{figure}

Exponential growth in scientific publications~\cite{bornmann2021growth} creates potential for cross-disciplinary insights that are largely untapped due to the limitations of manual literature review. Automating information extraction from this vast and varied body of work using AI is crucial to accelerate scientific progress. AI methods can be employed to identify and synthesize key findings, methodologies, and datasets across fields and thus to identify connections and facilitate the cross-pollination of ideas that would otherwise go unnoticed~\cite{buchkremer2019application,sourati2023accelerating}. 

Agentic workflows that leverage LLMs present a transformative new approach to engage with scientific literature. Employing autonomous agents with specific roles and capabilities makes it possible to automate the extraction of information and generation of structured datasets that represent key concepts and findings. Such datasets can be used to fine-tune models and enhance their ability to understand scientific text, answer domain-specific queries, and potentially contribute to tasks like hypothesis generation or literature summarization.

To explore the potential of agentic workflows for thus analyzing the scientific literature, we used \system to implement a system for generating and validating multi-choice questions (MCQs) from research publications~\cite{ch2018automatic,auroragpt-mcq}.
The workflow includes a PDFParser agent to extract text from a manuscript; two Generator agents that use different LLMs to generate MCQs; an MCQSelector to choose subsets of questions to evaluate; and two MCQAnswerers and two AnswerScorers (again, each with a different LLM) to generate and validate, respectively, answers to questions. The agents use the Mistral-7B-Instruct-v0.3~\cite{jiang2023mistral7b} and Meta-Llama-3-70B-Instruct~\cite{grattafiori2024llama3} models, denoted A and B, respectively. 
 
The beauty of this architecture is that alternative tasks and LLMs are easily integrated by defining new agents; agents can scale up and down in response to demand; and different agents can run concurrently or at different times. 
We show in \autoref{fig:mcq} an execution trace from a run in which the agents just listed were run concurrently to generate and validate MCQs for 10 publications. 

\section{Discussion}
\label{sec:discussion}
We describe several lessons learned implementing \system{} and applying it to a diverse range of applications.

\textbf{Scientific agentic systems require federation and statefulness.}
Science is often collaborative and multi-institutional. Research infrastructure (e.g., instruments, HPC systems, etc.) mirrors this inherently federated ecosystem. Building intelligent scientific systems requires agents that run in multiple locations and collaborate across institutions. 
Existing workflow systems (e.g., Parsl, Ray, Dask, Globus Compute) either do not support the capabilities necessary for agentic science, or face significant barriers when deployed on research infrastructure. 
Similarly, conversational agentic frameworks (e.g., LangChain, PydanticAI) are not designed for such environments, but can leverage \system{} to deliver these capabilities.

\textbf{Start-up barriers hinder adoption and deployment.} \system{} initially required that users deploy the hybrid exchange to allow inter-site communication. This requirement hindered the transition from local development to multi-site deployment. The introduction of the cloud exchange reduced barriers and thus enabled significant improvements (e.g., agent sharing and reuse). It also simplified integration with existing execution and data transfer solutions (e.g., Globus Compute, ProxyStore), further reducing barriers of using \system{}.

\textbf{Agentic middleware should not bias the behaviors that users create.} Commonly used agentic systems such as LangChain and SmolAgents shoehorn applications into specific patterns (e.g., LLMs with tool-calling).
Scientific agents behave in diverse ways that often do not fit such patterns. 
Rather than design specific behaviors into the system, we implemented patterns in \system{} (e.g., agent hierarchies, agents as resource managers) and reused them in different contexts.

\textbf{LLMs combined with tool calling cannot currently construct complex real-world science applications.} For instance, in the MOFA workflow we experimented with the use of LLMs for such functions as improving MOF design and deciding which simulators and simulations to run.
However, we found that the large number of tools and tool parameters, often with limited documentation, and sometimes subtle differences among tools, led to frequent problems including improperly configured/invoked codes that ran without errors but produced incorrect results. Manual construction of both tools (as \system{} actions) and workflow architecture (by invoking handles) informed by domain expertise, created more reliable applications.

\textbf{Autonomy and adaptability can improve application performance and resource utilization.} Static resource allocations often inadequately match the dynamic nature of scientific workloads. Agentic systems enable not only component autonomy but also dynamic resource utilization at fine granularity.

\section{Related Work}
\label{sec:related}

A \textbf{workflow}
is a structured sequence of tasks, typically a directed acyclic graph (DAG), designed to achieve a specific goal, often involving data transformation, analysis, or computational modeling.
Frameworks for building workflows take many forms.
Parallel computing libraries, such as Dask~\cite{rocklin2015dask} and Ray~\cite{moritz2018ray}, provide mechanisms for executing functions in parallel across local resources or distributed systems.
Similarly, workflow management systems (WMSs) can execute tasks in parallel but also provide mechanisms for defining, optimizing, and monitoring DAG execution (e.g., Airflow~\cite{airflow}, Fireworks~\cite{jain2015fireworks}, Makeflow~\cite{albrecht2012makeflow}, Nextflow~\cite{di2017nextflow}, Parsl~\cite{babuji2019parsl}, Pegasus~\cite{deelman15pegasus}, Swift~\cite{wilde2009parallel}).
WMSs can be differentiated by whether dependency graphs are defined~\cite{pauloski2024taps} with
static configurations files, such as CWL~\cite{crusoe2022cwl}, XML, or YAML; general purpose languages (GPLs); domain specific languages (DSLs); or procedurally through the dynamic execution of a program.
The class of workflows supported by these frameworks have two key limitations that we address: tasks are assumed to be pure (i.e., no side-effects) and programs are static, i.e., they cannot adapt to changing environments over time.

\textbf{Actors} are computational entities that enable concurrent computing through message passing~\cite{hewitt1973actors}.
In response to a message, an actor can alter its local state, send messages to other actors, and create new actors.
No global state means locks and synchronization primitives are not required.
Actors can enable stateful computations within traditionally stateless programming models, and are supported in parallel computing frameworks (e.g., Akka~\cite{akka}, Dask, Orleans~\cite{bernstein2014orleans}, Ray) and function-as-a-service (FaaS) platforms (e.g., Abaco~\cite{garcia2020abaco}, Azure Service Fabric~\cite{azureactors}, PraaS~\cite{copik2022praas}).
Actor models have been investigated as alternatives for designing computational workflows where communication and coordination are decoupled~\cite{bowers2005actor}.
Our system extends the actor model to support autonomous behaviors and federated deployments.

\textbf{Multi-agent systems} can enhance or automate scientific processes.
Early work investigated cooperative agent environments for distributed problem solving with minimal human intervention~\cite{drashansky1995sciagents,drashansky1999networked}.
Recent work focuses on improving the reasoning capabilities of LLM-backed agents through ontological knowledge graphs and multi-agent systems~\cite{ghafarollahi2024sciagents} and tool-augmented LLMs~\cite{ma2024sciagent}.
Increasingly popular is the use of multi-agent conversations, in which multiple role-specialized agents interact to collaborate, coordinate, or compete towards goals~\cite{wu2023autogen}.
These systems enhance LLM-based tools through better reasoning~\cite{du2023improving}, validation~\cite{wu2023agents}, and divergent thinking~\cite{liang2024encouraging}, prompting rapid development of frameworks such as LangGraph~\cite{langgraph}, Microsoft AutoGen~\cite{wu2023autogen}, OpenAI Swarm~\cite{openai-swarm}, and Pydantic Agents~\cite{pydantic-agents}.
Subsequently, interest in standardizing agent protocols has developed.
Anthropic's Model Context Protocol (MCP)~\cite{modelcontextprotocol} defines structured interaction between humans/tools and AI models.
Google's Agent2Agent (A2A) Protocol~\cite{agent2agentprotocol} focuses on structured interaction between autonomous agents; each agent serves an HTTP endpoint which is impracticable for many scientific workflows.
Multi-agent conversations can proxy scientists in iterative scientific processes---brainstorming ideas, planning experiments, and reasoning about results~\cite{boiko2023autonomous,wang2023scientific,gao2024agents,google2025coscientist}---but these aforementioned systems are designed for local or cloud-native applications and lack the features necessary to deploy agents across federated research infrastructure.
We focus on the systems-level challenges of representing and deploying diverse agent types and agentic workflows across heterogeneous environments rather than the applied use of LLMs for workflow steering.

\section{Conclusion \& Future Work}
Advancements in AI, coupled with concurrent advancements in self-driving laboratories, high performance computing, and research data management, open the door for truly autonomous scientific discovery.
Realizing this grand vision requires mechanisms for the seamless and dynamic integration of research software and infrastructure.
To that end, we introduced \system{}, a middleware for developing agentic workflows that engage multi-agent systems spanning federated research infrastructure.
This framework enables scalable and flexible orchestration of intelligent agents across heterogeneous resources.
We presented solutions to three key challenges: representing and programming agents; communicating among agents; and executing agents across diverse resources.
Our evaluations demonstrate that \system{} can support high-performance workflows, and four case studies highlight the advantages of agentic workflow design.

\section{Acknowledgments}
This research was supported in part by the National Science Foundation under Grants 2004894 and 2209919.
We used resources provided by the Argonne Leadership Computing Facility (ALCF), a U.S. Department of Energy (DOE) Office of Science user facility at Argonne National Laboratory supported under Contract DE-AC02-06CH11357, and the Chameleon testbed supported by the National Science Foundation.

\balance
\bibliographystyle{IEEEtran}
\bibliography{main}

\begin{thebibliography}{10}
\providecommand{\url}[1]{#1}
\csname url@samestyle\endcsname
\providecommand{\newblock}{\relax}
\providecommand{\bibinfo}[2]{#2}
\providecommand{\BIBentrySTDinterwordspacing}{\spaceskip=0pt\relax}
\providecommand{\BIBentryALTinterwordstretchfactor}{4}
\providecommand{\BIBentryALTinterwordspacing}{\spaceskip=\fontdimen2\font plus
\BIBentryALTinterwordstretchfactor\fontdimen3\font minus \fontdimen4\font\relax}
\providecommand{\BIBforeignlanguage}[2]{{%
\expandafter\ifx\csname l@#1\endcsname\relax
\typeout{** WARNING: IEEEtran.bst: No hyphenation pattern has been}%
\typeout{** loaded for the language `#1'. Using the pattern for}%
\typeout{** the default language instead.}%
\else
\language=\csname l@#1\endcsname
\fi
#2}}
\providecommand{\BIBdecl}{\relax}
\BIBdecl

\bibitem{zvyagin2023genslms}
M.~Zvyagin, A.~Brace, K.~Hippe, Y.~Deng, B.~Zhang, C.~O. Bohorquez, A.~Clyde, B.~Kale, D.~Perez-Rivera, H.~Ma \emph{et~al.}, ``{GenSLMs}: Genome-scale language models reveal {SARS}-{CoV}-2 evolutionary dynamics,'' \emph{The International Journal of High Performance Computing Applications}, vol.~37, no.~6, pp. 683--705, 2023.

\bibitem{deelman2009workflows}
E.~Deelman, D.~Gannon, M.~Shields, and I.~Taylor, ``Workflows and e-{S}cience: An overview of workflow system features and capabilities,'' \emph{Future Generation Computer Systems}, vol.~25, no.~5, pp. 528--540, 2009.

\bibitem{bryce2012saasglobus}
\BIBentryALTinterwordspacing
B.~Allen, J.~Bresnahan, L.~Childers, I.~Foster, G.~Kandaswamy, R.~Kettimuthu, J.~Kordas, M.~Link, S.~Martin, K.~Pickett, and S.~Tuecke, ``Software as a service for data scientists,'' \emph{Communications of the ACM}, vol.~55, no.~2, p. 81–88, feb 2012. [Online]. Available: \url{https://doi.org/10.1145/2076450.2076468}
\BIBentrySTDinterwordspacing

\bibitem{abolhasani2023sdl}
M.~Abolhasani and E.~Kumacheva, ``The rise of self-driving labs in chemical and materials sciences,'' \emph{Nature Synthesis}, vol.~2, no.~6, pp. 483--492, 2023.

\bibitem{pauloski2025agents}
\BIBentryALTinterwordspacing
J.~G. Pauloski, K.~Chard, and I.~Foster, ``Agentic discovery: Closing the loop with cooperative agents,'' \emph{Computer}, vol.~58, no.~10, pp. 20--27, October 2025. [Online]. Available: \url{https://doi.ieeecomputersociety.org/10.1109/MC.2025.3575029}
\BIBentrySTDinterwordspacing

\bibitem{wei2022chain}
J.~Wei, X.~Wang, D.~Schuurmans, M.~Bosma, F.~Xia, E.~Chi, Q.~V. Le, D.~Zhou \emph{et~al.}, ``Chain-of-thought prompting elicits reasoning in large language models,'' \emph{Advances in neural information processing systems}, vol.~35, pp. 24\,824--24\,837, 2022.

\bibitem{wang2022self}
X.~Wang, J.~Wei, D.~Schuurmans, Q.~Le, E.~Chi, S.~Narang, A.~Chowdhery, and D.~Zhou, ``Self-consistency improves chain of thought reasoning in language models,'' \emph{arXiv preprint arXiv:2203.11171}, 2022.

\bibitem{sakarvadia2024towards}
M.~Sakarvadia, ``Towards interpreting language models: A case study in multi-hop reasoning,'' \emph{arXiv preprint arXiv:2411.05037}, 2024.

\bibitem{wang2023scientific}
\BIBentryALTinterwordspacing
H.~Wang, T.~Fu, Y.~Du, W.~Gao, K.~Huang, Z.~Liu, P.~Chandak, S.~Liu, P.~V. Katwyk, A.~Deac, A.~Anandkumar, K.~J. Bergen, C.~P. Gomes, S.~Ho, P.~Kohli, J.~Lasenby, J.~Leskovec, T.-Y. Liu, A.~K. Manrai, D.~S. Marks, B.~Ramsundar, L.~Song, J.~Sun, J.~Tang, P.~Velickovic, M.~Welling, L.~Zhang, C.~W. Coley, Y.~Bengio, and M.~Zitnik, ``Scientific discovery in the age of artificial intelligence,'' \emph{Nature}, vol. 620, pp. 47--60, 2023. [Online]. Available: \url{https://api.semanticscholar.org/CorpusID:260384616}
\BIBentrySTDinterwordspacing

\bibitem{gao2024agents}
\BIBentryALTinterwordspacing
S.~Gao, A.~Fang, Y.~Huang, V.~Giunchiglia, A.~Noori, J.~R. Schwarz, Y.~Ektefaie, J.~Kondic, and M.~Zitnik, ``Empowering biomedical discovery with {AI} agents,'' \emph{Cell}, vol. 187, no.~22, pp. 6125--6151, 2024. [Online]. Available: \url{https://www.sciencedirect.com/science/article/pii/S0092867424010705}
\BIBentrySTDinterwordspacing

\bibitem{wu2023autogen}
\BIBentryALTinterwordspacing
Q.~Wu, G.~Bansal, J.~Zhang, Y.~Wu, B.~Li, E.~Zhu, L.~Jiang, X.~Zhang, S.~Zhang, J.~Liu, A.~H. Awadallah, R.~W. White, D.~Burger, and C.~Wang, ``{AutoGen}: Enabling next-gen {LLM} applications via multi-agent conversation,'' 2023. [Online]. Available: \url{https://arxiv.org/abs/2308.08155}
\BIBentrySTDinterwordspacing

\bibitem{langgraph}
\BIBentryALTinterwordspacing
LangChain, ``{LangGraph},'' 2024. [Online]. Available: \url{https://www.langchain.com/langgraph}
\BIBentrySTDinterwordspacing

\bibitem{openai-swarm}
\BIBentryALTinterwordspacing
OpenAI, ``Swarm,'' 2024. [Online]. Available: \url{https://github.com/openai/swarm}
\BIBentrySTDinterwordspacing

\bibitem{osti2023iri}
\BIBentryALTinterwordspacing
W.~L. Miller, D.~Bard, A.~Boehnlein, K.~Fagnan, C.~Guok, E.~Lançon, S.~Ramprakash, M.~Shankar, N.~Schwarz, and B.~L. Brown, ``Integrated {R}esearch {I}nfrastructure {A}rchitecture {B}lueprint {A}ctivity ({F}inal {R}eport 2023),'' US Department of Energy (USDOE), Washington, DC (United States). Office of Science; Lawrence Berkeley National Laboratory (LBNL), Berkeley, CA (United States), Tech. Rep., 07 2023. [Online]. Available: \url{https://www.osti.gov/biblio/1984466}
\BIBentrySTDinterwordspacing

\bibitem{chard2020funcx}
\BIBentryALTinterwordspacing
R.~Chard, Y.~Babuji, Z.~Li, T.~Skluzacek, A.~Woodard, B.~Blaiszik, I.~Foster, and K.~Chard, ``{funcX}: A federated function serving fabric for science,'' in \emph{29th International Symposium on High-Performance Parallel and Distributed Computing}, ser. HPDC '20.\hskip 1em plus 0.5em minus 0.4em\relax New York, NY, USA: Association for Computing Machinery, 2020, p. 65–76. [Online]. Available: \url{https://doi.org/10.1145/3369583.3392683}
\BIBentrySTDinterwordspacing

\bibitem{goodwin1995formalizing}
R.~Goodwin, ``Formalizing properties of agents,'' \emph{Journal of Logic and Computation}, vol.~5, no.~6, pp. 763--781, 1995.

\bibitem{wooldridge1995agents}
\BIBentryALTinterwordspacing
M.~Wooldridge and N.~R. Jennings, ``Intelligent agents: Theory and practice,'' \emph{The Knowledge Engineering Review}, vol.~10, pp. 115 -- 152, 1995. [Online]. Available: \url{https://api.semanticscholar.org/CorpusID:221342993}
\BIBentrySTDinterwordspacing

\bibitem{nwana1996agents}
H.~S. Nwana, ``Software agents: An overview,'' \emph{The Knowledge Engineering Review}, vol.~11, no.~3, p. 205–244, 1996.

\bibitem{panait2005cooperative}
L.~Panait and S.~Luke, ``Cooperative multi-agent learning: The state of the art,'' \emph{Autonomous agents and multi-agent systems}, vol.~11, pp. 387--434, 2005.

\bibitem{hewitt1973actors}
C.~Hewitt, P.~Bishop, and R.~Steiger, ``A universal modular {ACTOR} formalism for artificial intelligence,'' in \emph{3rd International Joint Conference on Artificial Intelligence}, ser. IJCAI'73.\hskip 1em plus 0.5em minus 0.4em\relax San Francisco, CA, USA: Morgan Kaufmann Publishers Inc., 1973, p. 235–245.

\bibitem{ward2021colmena}
L.~Ward, G.~Sivaraman, J.~G. Pauloski, Y.~Babuji, R.~Chard, N.~Dandu, P.~C. Redfern, R.~S. Assary, K.~Chard, L.~A. Curtiss, R.~Thakur, and I.~Foster, ``Colmena: Scalable machine-learning-based steering of ensemble simulations for high performance computing,'' in \emph{IEEE/ACM Workshop on Machine Learning in High Performance Computing Environments}.\hskip 1em plus 0.5em minus 0.4em\relax IEEE, 2021, pp. 9--20.

\bibitem{kolodisner2025viral}
N.~Kolodisner, A.~Kamatar, and J.~G. Pauloski, ``An agent-based viral venture: Adaptive tool selection for scalable genomics,'' in \emph{International Conference for High Performance Computing, Networking, Storage and Analysis - SRC and Research Posters (In Press)}, 2025.

\bibitem{tynes2025blend}
M.~Tynes, K.~Chard, I.~Foster, and L.~Ward, ``Will it blend? mixing numerical and machine-learned physics quantities for accurate on-the-fly surrogate modeling,'' in \emph{Computational Science -- ICCS 2025}, M.~H. Lees, W.~Cai, S.~A. Cheong, Y.~Su, D.~Abramson, J.~J. Dongarra, and P.~M.~A. Sloot, Eds.\hskip 1em plus 0.5em minus 0.4em\relax Cham: Springer Nature Switzerland, 2025, pp. 270--284.

\bibitem{yan25mofa}
\BIBentryALTinterwordspacing
X.~Yan, N.~Hudson, H.~Park, D.~Grzenda, J.~G. Pauloski, M.~Schwarting, H.~Pan, H.~Harb, S.~Foreman, C.~Knight, T.~Gibbs, K.~Chard, S.~Chaudhuri, E.~Tajkhorshid, I.~Foster, M.~Moosavi, L.~Ward, and E.~A. Huerta, ``{MOFA}: Discovering materials for carbon capture with a {GenAI}- and simulation-based workflow,'' 2025. [Online]. Available: \url{https://arxiv.org/abs/2501.10651}
\BIBentrySTDinterwordspacing

\bibitem{saville2022cepi}
\BIBentryALTinterwordspacing
M.~Saville, J.~P. Cramer, M.~Downham, A.~Hacker, N.~Lurie, L.~V. der Veken, M.~Whelan, and R.~Hatchett, ``Delivering pandemic vaccines in 100 days — what will it take?'' \emph{New England Journal of Medicine}, vol. 387, no.~2, p.~e3, 2022. [Online]. Available: \url{https://www.nejm.org/doi/full/10.1056/NEJMp2202669}
\BIBentrySTDinterwordspacing

\bibitem{vescovi2022linking}
R.~Vescovi, R.~Chard, N.~Saint, B.~Blaisik, J.~Pruyne, T.~Bicer, A.~Lavens, Z.~Liu, M.~E. Papka, S.~Narayanan, N.~Schwarz, K.~Chard, and I.~Foster, ``Linking scientific instruments and computation: Patterns, technologies and experiences,'' \emph{Patterns}, vol.~3, no.~10, 2022.

\bibitem{babnigg2025ifum}
D.~Babnigg, ``Unraveling distant galaxies: Analyzing {IFU} data with {P}arsl and {A}cademy,'' in \emph{International Conference for High Performance Computing, Networking, Storage and Analysis - SRC and Research Posters}, 2025.

\bibitem{vescovi2023sdl}
R.~Vescovi, T.~Ginsburg, K.~Hippe, D.~Ozgulbas, C.~Stone, A.~Stroka, R.~Butler, B.~Blaiszik, T.~Brettin, K.~Chard \emph{et~al.}, ``Towards a modular architecture for science factories,'' \emph{Digital Discovery}, vol.~2, no.~6, pp. 1980--1998, 2023.

\bibitem{coburn2025drmacs}
J.~Coburn, A.~Wells, N.~Ramachandra, and S.~Habib, ``A multiagent system for cosmological data analysis,'' \emph{Submitted for publication to AAMAS-2026}, 2025.

\bibitem{pham2025chemgraph}
T.~D. Pham, A.~Tanikanti, and M.~Ke{\c{c}}eli, ``Chem{G}raph: An agentic framework for computational chemistry workflows,'' \emph{arXiv preprint arXiv:2506.06363}, 2025.

\bibitem{papka2024alcfops}
\BIBentryALTinterwordspacing
M.~E. Papka, W.~Allcock, B.~Cerny, J.~Francis, K.~Kumaran, A.~Madduri, A.~L. Manning, V.~Mateevitsi, J.~Neel, A.~Pope \emph{et~al.}, ``2024 operational assessment report - argonne leadership computing facility,'' Argonne National Laboratory (ANL), Tech. Rep., 12 2024. [Online]. Available: \url{https://www.osti.gov/biblio/2574032}
\BIBentrySTDinterwordspacing

\bibitem{babuji2019parsl}
\BIBentryALTinterwordspacing
Y.~Babuji, A.~Woodard, Z.~Li, D.~S. Katz, B.~Clifford, R.~Kumar, L.~Lacinski, R.~Chard, J.~M. Wozniak, I.~Foster, M.~Wilde, and K.~Chard, ``Parsl: Pervasive parallel programming in {P}ython,'' in \emph{28th International Symposium on High-Performance Parallel and Distributed Computing}, ser. HPDC '19.\hskip 1em plus 0.5em minus 0.4em\relax New York, NY, USA: Association for Computing Machinery, 2019, p. 25–36. [Online]. Available: \url{https://doi.org/10.1145/3307681.3325400}
\BIBentrySTDinterwordspacing

\bibitem{tuecke2016auth}
S.~Tuecke, R.~Ananthakrishnan, K.~Chard, M.~Lidman, B.~McCollam, S.~Rosen, and I.~Foster, ``Globus auth: A research identity and access management platform,'' in \emph{IEEE 12th International Conference on e-Science}, 2016, pp. 203--212.

\bibitem{pauloski2023proxystore}
\BIBentryALTinterwordspacing
J.~G. Pauloski, V.~Hayot-Sasson, L.~Ward, N.~Hudson, C.~Sabino, M.~Baughman, K.~Chard, and I.~Foster, ``Accelerating communications in federated applications with transparent object proxies,'' in \emph{International Conference for High Performance Computing, Networking, Storage and Analysis}, ser. SC '23.\hskip 1em plus 0.5em minus 0.4em\relax New York, NY, USA: ACM, 2023. [Online]. Available: \url{https://doi.org/10.1145/3581784.3607047}
\BIBentrySTDinterwordspacing

\bibitem{pauloski2024proxystore}
J.~G. Pauloski, V.~Hayot-Sasson, L.~Ward, A.~Brace, A.~Bauer, K.~Chard, and I.~Foster, ``Object proxy patterns for accelerating distributed applications,'' \emph{IEEE Transactions on Parallel and Distributed Systems}, vol.~36, no.~2, pp. 253--265, 2025.

\bibitem{ross2020mochi}
R.~Ross, G.~Amvrosiadis, P.~Carns, C.~D. Cranor, M.~Dorier, K.~Harms, G.~Ganger, G.~Gibson, S.~Gutierrez, R.~Latham, B.~Robey, D.~Robinson, B.~Settlemyer, G.~Shipman, S.~Snyder, J.~Soumagne, and Z.~Qing, ``Mochi: Composing data services for high-performance computing environments,'' \emph{Journal of Computer Science and Technology}, vol.~35, no.~1, pp. 121 -- 144,, Jan 2020, 10.1007/s11390-020-9802-0.

\bibitem{shamis2015ucx}
P.~Shamis, M.~G. Venkata, M.~G. Lopez, M.~B. Baker, O.~Hernandez, Y.~Itigin, M.~Dubman, G.~Shainer, R.~L. Graham, L.~Liss, Y.~Shahar, S.~Potluri, D.~Rossetti, D.~Becker, D.~Poole, C.~Lamb, S.~Kumar, C.~Stunkel, G.~Bosilca, and A.~Bouteiller, ``{UCX}: An open source framework for {HPC} network {APIs} and beyond,'' in \emph{IEEE 23rd Annual Symposium on High-Performance Interconnects}.\hskip 1em plus 0.5em minus 0.4em\relax IEEE, 2015, pp. 40--43.

\bibitem{chard2014globus}
K.~Chard, S.~Tuecke, and I.~Foster, ``Efficient and secure transfer, synchronization, and sharing of big data,'' \emph{IEEE Cloud Computing}, vol.~1, no.~3, pp. 46--55, 2014.

\bibitem{keahey2020lessons}
K.~Keahey, J.~Anderson, Z.~Zhen, P.~Riteau, P.~Ruth, D.~Stanzione, M.~Cevik, J.~Colleran, H.~S. Gunawi, C.~Hammock, J.~Mambretti, A.~Barnes, F.~Halbach, A.~Rocha, and J.~Stubbs, ``Lessons learned from the {C}hameleon testbed,'' in \emph{USENIX Annual Technical Conference}.\hskip 1em plus 0.5em minus 0.4em\relax USENIX Association, July 2020.

\bibitem{furukawa2013chemistry}
H.~Furukawa, K.~E. Cordova, M.~O’Keeffe, and O.~M. Yaghi, ``The chemistry and applications of metal-organic frameworks,'' \emph{Science}, vol. 341, no. 6149, p. 1230444, 2013.

\bibitem{zhao2025artificial}
Y.~Zhao, Y.~Zhao, J.~Wang, and Z.~Wang, ``Artificial intelligence meets laboratory automation in discovery and synthesis of metal--organic frameworks: A review,'' \emph{Industrial \& Engineering Chemistry Research}, vol.~64, no.~9, pp. 4637--4668, 2025.

\bibitem{mateo2022ifum}
\BIBentryALTinterwordspacing
M.~Mateo, J.~I.~B. III, Y.~Song, J.~Crane, C.~Hull, S.~Shectman, and C.~Birk, ``{IFUM}: Integral field units for {M}agellan,'' in \emph{Ground-based and Airborne Instrumentation for Astronomy IX}, C.~J. Evans, J.~J. Bryant, and K.~Motohara, Eds., vol. 12184, International Society for Optics and Photonics.\hskip 1em plus 0.5em minus 0.4em\relax SPIE, 2022, p. 121845P. [Online]. Available: \url{https://doi.org/10.1117/12.2629506}
\BIBentrySTDinterwordspacing

\bibitem{hegedHus2019gossip}
I.~Heged{\H{u}}s, G.~Danner, and M.~Jelasity, ``Gossip learning as a decentralized alternative to federated learning,'' in \emph{Distributed Applications and Interoperable Systems: 19th IFIP WG 6.1 International Conference, DAIS 2019, Held as Part of the 14th International Federated Conference on Distributed Computing Techniques, DisCoTec 2019, Kongens Lyngby, Denmark, June 17--21, 2019, Proceedings 19}.\hskip 1em plus 0.5em minus 0.4em\relax Springer, 2019, pp. 74--90.

\bibitem{holme2002growing}
P.~Holme and B.~J. Kim, ``Growing scale-free networks with tunable clustering,'' \emph{Physical Review E}, vol.~65, no.~2, p. 026107, 2002.

\bibitem{deng2012mnist}
L.~Deng, ``The {MNIST} database of handwritten digit images for machine learning research,'' \emph{IEEE Signal Processing Magazine}, vol.~29, no.~6, pp. 141--142, 2012.

\bibitem{bornmann2021growth}
L.~Bornmann, R.~Haunschild, and R.~Mutz, ``Growth rates of modern science: A latent piecewise growth curve approach to model publication numbers from established and new literature databases,'' \emph{Humanities and Social Sciences Communications}, vol.~8, no.~1, pp. 1--15, 2021.

\bibitem{buchkremer2019application}
R.~Buchkremer, A.~Demund, S.~Ebener, F.~Gampfer, D.~J{\"a}gering, A.~J{\"u}rgens, S.~Klenke, D.~Krimpmann, J.~Schmank, M.~Spiekermann, M.~Wahlers, and M.~Wiepke, ``The application of artificial intelligence technologies as a substitute for reading and to support and enhance the authoring of scientific review articles,'' \emph{IEEE access}, vol.~7, pp. 65\,263--65\,276, 2019.

\bibitem{sourati2023accelerating}
J.~Sourati and J.~A. Evans, ``Accelerating science with human-aware artificial intelligence,'' \emph{Nature Human Behaviour}, vol.~7, no.~10, pp. 1682--1696, 2023.

\bibitem{ch2018automatic}
D.~R. Ch and S.~K. Saha, ``Automatic multiple choice question generation from text: A survey,'' \emph{IEEE Transactions on Learning Technologies}, vol.~13, no.~1, pp. 14--25, 2018.

\bibitem{auroragpt-mcq}
\BIBentryALTinterwordspacing
C.~Catlett and I.~Foster, ``Creating and scoring multiple choice questions (mcqs) from papers,'' 2025. [Online]. Available: \url{https://github.com/auroraGPT-ANL/MCQ-and-SFT-code}
\BIBentrySTDinterwordspacing

\bibitem{jiang2023mistral7b}
\BIBentryALTinterwordspacing
A.~Q. Jiang, A.~Sablayrolles, A.~Mensch, C.~Bamford, D.~S. Chaplot, D.~de~las Casas, F.~Bressand, G.~Lengyel, G.~Lample, L.~Saulnier, L.~R. Lavaud, M.-A. Lachaux, P.~Stock, T.~L. Scao, T.~Lavril, T.~Wang, T.~Lacroix, and W.~E. Sayed, ``Mistral {7B},'' 2023. [Online]. Available: \url{https://arxiv.org/abs/2310.06825}
\BIBentrySTDinterwordspacing

\bibitem{grattafiori2024llama3}
\BIBentryALTinterwordspacing
T.~L.~T. at~Meta, ``The {L}lama 3 herd of models,'' 2024. [Online]. Available: \url{https://arxiv.org/abs/2407.21783}
\BIBentrySTDinterwordspacing

\bibitem{rocklin2015dask}
M.~Rocklin, ``Dask: Parallel computation with blocked algorithms and task scheduling,'' in \emph{14th Python in Science Conference}, K.~Huff and J.~Bergstra, Eds.\hskip 1em plus 0.5em minus 0.4em\relax Austin, TX, USA: SciPy, 2015, pp. 126 -- 132.

\bibitem{moritz2018ray}
P.~Moritz, R.~Nishihara, S.~Wang, A.~Tumanov, R.~Liaw, E.~Liang, M.~Elibol, Z.~Yang, W.~Paul, M.~I. Jordan, and I.~Stoica, ``Ray: A distributed framework for emerging {AI} applications,'' in \emph{13th USENIX Conference on Operating Systems Design and Implementation}, ser. OSDI'18.\hskip 1em plus 0.5em minus 0.4em\relax USA: USENIX Association, 2018, p. 561–577.

\bibitem{airflow}
\BIBentryALTinterwordspacing
Apache, ``Airflow,'' 2015. [Online]. Available: \url{https://airflow.apache.org/}
\BIBentrySTDinterwordspacing

\bibitem{jain2015fireworks}
A.~Jain, S.~P. Ong, W.~Chen, B.~Medasani, X.~Qu, M.~Kocher, M.~Brafman, G.~Petretto, G.-M. Rignanese, G.~Hautier, D.~Gunter, and K.~A. Persson, ``{FireWorks}: A dynamic workflow system designed for high-throughput applications,'' \emph{Concurrency and Computation: Practice and Experience}, vol.~27, no.~17, pp. 5037--5059, 2015.

\bibitem{albrecht2012makeflow}
M.~Albrecht, P.~Donnelly, P.~Bui, and D.~Thain, ``Makeflow: A portable abstraction for data intensive computing on clusters, clouds, and grids,'' in \emph{1st ACM SIGMOD Workshop on Scalable Workflow Execution Engines and Technologies}, ser. SWEET '12.\hskip 1em plus 0.5em minus 0.4em\relax New York, NY, USA: Association for Computing Machinery, 2012.

\bibitem{di2017nextflow}
P.~Di~Tommaso, M.~Chatzou, E.~W. Floden, P.~P. Barja, E.~Palumbo, and C.~Notredame, ``Nextflow enables reproducible computational workflows,'' \emph{Nature Biotechnology}, vol.~35, no.~4, pp. 316--319, 2017.

\bibitem{deelman15pegasus}
E.~Deelman, K.~Vahi, G.~Juve, M.~Rynge, S.~Callaghan, P.~J. Maechling, R.~Mayani, W.~Chen, R.~{Ferreira da Silva}, M.~Livny, and K.~Wenger, ``Pegasus, a workflow management system for science automation,'' \emph{Future Generation Computer Systems}, vol.~46, pp. 17--35, 2015.

\bibitem{wilde2009parallel}
M.~Wilde, I.~Foster, K.~Iskra, P.~Beckman, Z.~Zhang, A.~Espinosa, M.~Hategan, B.~Clifford, and I.~Raicu, ``Parallel scripting for applications at the petascale and beyond,'' \emph{Computer}, vol.~42, no.~11, pp. 50--60, 2009.

\bibitem{pauloski2024taps}
J.~G. Pauloski, V.~Hayot-Sasson, M.~Gonthier, N.~Hudson, H.~Pan, S.~Zhou, I.~Foster, and K.~Chard, ``{TaPS}: A performance evaluation suite for task-based execution frameworks,'' in \emph{IEEE 20th International Conference on e-Science}.\hskip 1em plus 0.5em minus 0.4em\relax New York, NY, USA: IEEE, 2024, pp. 1--10.

\bibitem{crusoe2022cwl}
\BIBentryALTinterwordspacing
M.~R. Crusoe, S.~Abeln, A.~Iosup, P.~Amstutz, J.~Chilton, N.~Tijani\'{c}, H.~M\'{e}nager, S.~Soiland-Reyes, B.~Gavrilovi\'{c}, C.~Goble, and T.~C. Community, ``Methods included: Standardizing computational reuse and portability with the {C}ommon {W}orkflow {L}anguage,'' \emph{Commun. ACM}, vol.~65, no.~6, p. 54–63, May 2022. [Online]. Available: \url{https://doi.org/10.1145/3486897}
\BIBentrySTDinterwordspacing

\bibitem{akka}
\BIBentryALTinterwordspacing
Lightbend, ``Akka: The actor model on the {JVM},'' 2009. [Online]. Available: \url{https://akka.io/}
\BIBentrySTDinterwordspacing

\bibitem{bernstein2014orleans}
P.~Bernstein, S.~Bykov, A.~Geller, G.~Kliot, and J.~Thelin, ``Orleans: Distributed virtual actors for programmability and scalability,'' Microsoft, Tech. Rep. MSR-TR-2014-41, March 2014.

\bibitem{garcia2020abaco}
C.~Garcia, J.~Stubbs, J.~Looney, A.~Jamthe, and M.~Packard, ``Abaco--{A} modern platform for high throughput parallel rcientific computations,'' 2020.

\bibitem{azureactors}
\BIBentryALTinterwordspacing
Microsoft, ``Azure: Service fabric reliable actors,'' 2017. [Online]. Available: \url{https://learn.microsoft.com/en-us/azure/service-fabric/service-fabric-reliable-actors-introduction}
\BIBentrySTDinterwordspacing

\bibitem{copik2022praas}
M.~Copik, A.~Calotoiu, R.~Bruno, G.~Rethy, R.~Böhringer, and T.~Hoefler, ``Process-as-a-service: Elastic and stateful serverless with cloud processes,'' ETH Zürich, Tech. Rep., 01 2022.

\bibitem{bowers2005actor}
S.~Bowers and B.~Lud{\"a}scher, ``Actor-oriented design of scientific workflows,'' in \emph{Conceptual Modeling -- ER 2005}, L.~Delcambre, C.~Kop, H.~C. Mayr, J.~Mylopoulos, and O.~Pastor, Eds.\hskip 1em plus 0.5em minus 0.4em\relax Berlin, Heidelberg: Springer Berlin Heidelberg, 2005, pp. 369--384.

\bibitem{drashansky1995sciagents}
T.~T. Drashansky, A.~Joshi, and J.~R. Rice, ``Sci{A}gents-an agent based environment for distributed, cooperative scientific computing,'' in \emph{7th IEEE International Conference on Tools with Artificial Intelligence}.\hskip 1em plus 0.5em minus 0.4em\relax IEEE, 1995, pp. 452--459.

\bibitem{drashansky1999networked}
T.~Drashansky, E.~N. Houstis, N.~Ramakrishnan, and J.~R. Rice, ``Networked agents for scientific computing,'' \emph{Communications of the ACM}, vol.~42, no.~3, pp. 48--ff, 1999.

\bibitem{ghafarollahi2024sciagents}
A.~Ghafarollahi and M.~J. Buehler, ``Sci{A}gents: Automating scientific discovery through bioinspired multi-agent intelligent graph reasoning,'' \emph{Advanced Materials}, p. 2413523, 2024.

\bibitem{ma2024sciagent}
Y.~Ma, Z.~Gou, J.~Hao, R.~Xu, S.~Wang, L.~Pan, Y.~Yang, Y.~Cao, A.~Sun, H.~Awadalla \emph{et~al.}, ``Sci{A}gent: Tool-augmented language models for scientific reasoning,'' \emph{arXiv preprint arXiv:2402.11451}, 2024.

\bibitem{du2023improving}
\BIBentryALTinterwordspacing
Y.~Du, S.~Li, A.~Torralba, J.~B. Tenenbaum, and I.~Mordatch, ``Improving factuality and reasoning in language models through multiagent debate,'' 2023. [Online]. Available: \url{https://arxiv.org/abs/2305.14325}
\BIBentrySTDinterwordspacing

\bibitem{wu2023agents}
\BIBentryALTinterwordspacing
Y.~Wu, F.~Jia, S.~Zhang, H.~Li, E.~Zhu, Y.~Wang, Y.~T. Lee, R.~Peng, Q.~Wu, and C.~Wang, ``{MathChat}: Converse to tackle challenging math problems with {LLM} agents,'' 2024. [Online]. Available: \url{https://arxiv.org/abs/2306.01337}
\BIBentrySTDinterwordspacing

\bibitem{liang2024encouraging}
\BIBentryALTinterwordspacing
T.~Liang, Z.~He, W.~Jiao, X.~Wang, Y.~Wang, R.~Wang, Y.~Yang, S.~Shi, and Z.~Tu, ``Encouraging divergent thinking in large language models through multi-agent debate,'' in \emph{2024 Conference on Empirical Methods in Natural Language Processing}, Y.~Al-Onaizan, M.~Bansal, and Y.-N. Chen, Eds.\hskip 1em plus 0.5em minus 0.4em\relax Miami, Florida, USA: Association for Computational Linguistics, Nov. 2024, pp. 17\,889--17\,904. [Online]. Available: \url{https://aclanthology.org/2024.emnlp-main.992/}
\BIBentrySTDinterwordspacing

\bibitem{pydantic-agents}
\BIBentryALTinterwordspacing
Pydantic, ``Agents,'' 2024. [Online]. Available: \url{https://ai.pydantic.dev/agents/}
\BIBentrySTDinterwordspacing

\bibitem{modelcontextprotocol}
\BIBentryALTinterwordspacing
Anthropic, ``{Model Context Protocol (MCP}),'' 2024. [Online]. Available: \url{https://modelcontextprotocol.io/}
\BIBentrySTDinterwordspacing

\bibitem{agent2agentprotocol}
\BIBentryALTinterwordspacing
Google, ``Agent2{A}gent {P}rotocol ({A2A}),'' 2025. [Online]. Available: \url{https://github.com/google/A2A}
\BIBentrySTDinterwordspacing

\bibitem{boiko2023autonomous}
D.~A. Boiko, R.~MacKnight, B.~Kline, and G.~Gomes, ``Autonomous chemical research with large language models,'' \emph{Nature}, vol. 624, no. 7992, pp. 570--578, 2023.

\bibitem{google2025coscientist}
\BIBentryALTinterwordspacing
{Google Research}, ``Accelerating scientific breakthroughs with an {AI} co-scientist,'' 2025. [Online]. Available: \url{https://research.google/blog/accelerating-scientific-breakthroughs-with-an-ai-co-scientist/}
\BIBentrySTDinterwordspacing

\end{thebibliography}

\end{document}